# Slow but complete, two state unfolding/refolding of lysozyme in "tuned" ~6M guanidinium carboxylate solutions.

Zuofeng Zhao and Austen Angell
Dept. of Chemistry and Biochemistry, Arizona State University, Tempe, AZ 85287-1604

**Abstract.**
We use differential scanning calorimetry of the unfolding process to identify conditions in which the pseudo two-state refolding of thermally denatured lysozyme can be observed to occur on time scales of hours, in solutions that approach 6M in guanidinium cation, Gdm$^+$. Remarkably, the fraction of lysozyme re-folded at 25ºC reaches, and remains at 1.0. The refolded fraction is linear in log(waiting time), where the waiting time is the time at ambient temperature after an initial thermal denaturing (details in text) and immediate rapid cool to ambient. The favorable refolding conditions are achieved by tuning the solution anion composition to be comparable in pKa to, but somewhat smaller than, the pKa values of the carboxylate residues, aspartic acid and glutamic acid in the heteropolymer chain. To date we have used mixtures of guanidinium formate and acetate in equal proportions, with sufficient water to achieve the Gdm$^+$ concentrations 5.36 and 4.39M. Reducing the concentration increases the folding rate without changing the final 100% refolded state. We suggest the slowdown occurs because we have chemically pre-empted the critical links that nucleate the folding process and determine its all-or-nothing character.

**Introduction.**

As is made clear by the continuing controversies on how guanidinium cations interact with proteins(*1*) - usually studied in relation to denaturing effects - the factors determining whether or not a protein in the unfolded state can refold reliably to its native state remain incompletely understood. While much progress has been made(*2-5*), particularly through the study of mutant versions (both natural(*6*) and engineered(*7, 8*)) of the natural sequence, and through the analysis of various statistical mechanical models(*9-12*) the precise sequence of events that permits the protein to almost unerringly regain its ground state after thermal or chemical unfolding (*13*), remains a matter for discussion. While it seems clear that, for small fast-folding proteins there are a variety of paths that can be followed(*14, 15*), for more complex cases the number of routes becomes increasingly restricted.

The targets of greatest interest, to the present authors, are the so-called "2-state folders" of which many are documented. For a long time lysozyme was considered to be among this simple class(*16, 17*), though it is now appreciated that, under physiological conditions at least, there is an intermediate state that might be encountered, and must be escaped from, along some routes, i.e. it is only a pseudo two state case. Lysozyme has been the object of more studies than almost any other protein, probably because of its abundance and ease of purification, and we will use it in the present study despite its diminished status.

It is intrinsic to their function that proteins, below their denaturation temperatures $T_d$, are in a state of dynamic equilibrium between folded and unfolded states, and have their maximum refolding rates near the physiological pH. If stability of a folded protein is defined in terms of the denaturation ("melting") temperature, then lysozyme is a rather stable protein with $T_m$ as high as 80ºC in physiological solutions where it is leveling off with increasing pH.  Privalov and coworkers do not report $T_m$ beyond pH 5. In hydrated ionic liquid media, a maximum in unfolding enthalpy can be observed(*18*) with increasing basicity but the $T_m$ is still increasing at the maximum basicity studied(*19*) (in hydrated dibutylammonium acetate where refoldability index is zero).  An alternative useful stability index refers to the ability to refold after denaturation, for which a refoldability (fraction, or index) can be defined.

Both Td and the refoldability are conveniently determined calorimetrically during temperature upscans at a constant rate. The melting temperature falls at the inflection point of the sigmoid defining the course of the temperature-induced conversion from folded state to the higher entropy unfolded state. Indeed it was the conformity to the demands of a simple two-state equilibrium that was used by Privalov and co-workers to decide whether a given unfolding protein could be classed as a 2-state folder or not. Lysozyme satisfies this criterion under both physiological conditions and in the high ionic strength conditions that this laboratory has been using in recent studies.  If the unfolding and refolding rates are high relative to the scan rate used in the study then the inflection point of the sigmoid will be determined entirely by the thermodynamic parameters, $\Delta H$ and $\Delta S$ governing the equilibrium.  On the other hand if we drastically decrease the refolding rate then the system will no longer follow the dictates of thermodynamics, and the folding phenomenon will take on new aspects.

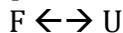

In past works, we have controlled the refolding rate by the strategy of fast quenching to preserve the unfolded state in a glassy matrix and then refolding at the rate of our choice(*20*). In the present work we control the refolding rate chemically by choosing the nature of the medium in which the protein is dissolved. This has previously been accomplished by reducing concentrations of conventional denaturants, often presented as Chevron plots(*4*).

In order to emphasize the importance of the anion effects we choose as cation the guanidinium ion that is normally associated with powerful denaturation effects, but is also known, since early work(*21*),  to be non-destabilizing,  when in combination with more basic anions.  To revisit this familiar scene, and to add to the usual anion retinue the case that will be of special importance in our study, we display the denaturation (or "melting") temperatures, determined using differential scanning calorimetry as before(*18*), in Figure 1.  The melting temperatures plotted are the peaks of unfolding  endotherms like those seen in Figure 2.  Guanidinium formate, the preparation of which is described in a recent paper on guanidinium salt physical properties(*22*), is seen to lower the melting temperature a little more rapidly than does Gdm acetate (which itself is much less water-soluble). We need to make a clear distinction between the thermal stability determined by the melting temperature, and the stability that is manifested by refolding propensity or resistance to aggregation since these can vary in opposite directions. Figure 1 shows that the thermal stability decreases with increasing concentration of Gdm formate more rapidly than in the case of Gdm acetate while, with Gdm sulfate, the melting temperature increases with increasing concentration as documented many times before(*21*). In hydrated ammonium acetate the $T_m$ rises as high as 85ºC but refolding does not occur unless some organic cation is included in the solution. Likewise we find that once unfolded in Gdm sulfate solutions, the refolding on cooling is uncertain at best, whereas, in the mixtures of Gdm acetate and formate that we now describe, the refolding of lysozyme, while slow, is remarkably complete.

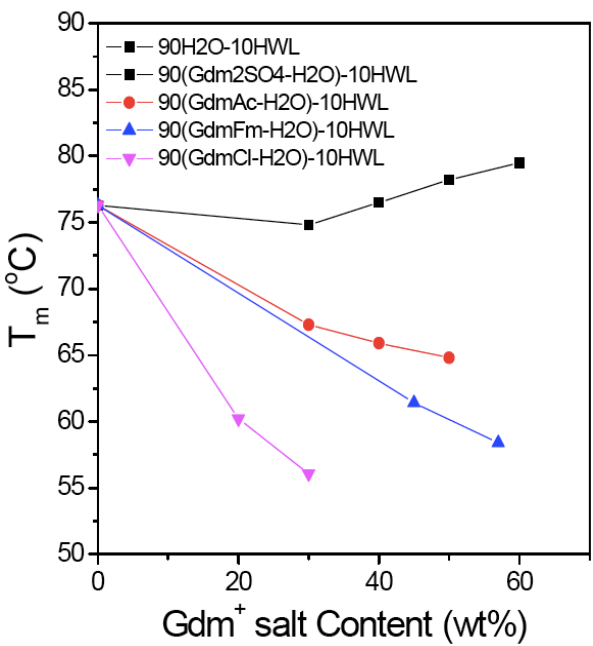

**Figure 1.** Denaturation ("melting") temperatures of lysozyme in aqueous solutions of the guanidinium salts indicated in the legend as function of salt content. The hen egg white lysozyme (HEWL) content of the final solutions is 10 wt % in each case.

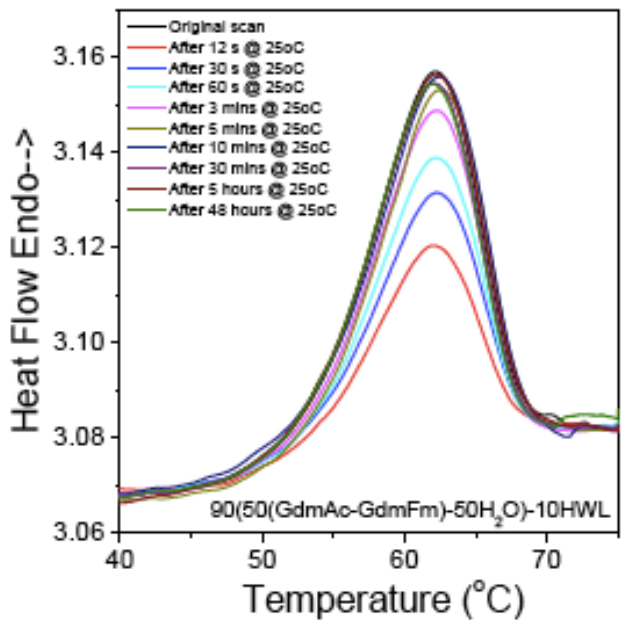

**Figure 2**. Unfolding endotherms for lysozyme in a GdmFm/GdmAc solution of Gdm+ concentration 4.39M. Dark curve is endotherm for the initial unfolding. Colored curves are for the same sample after holding at ambient temperature for the times given in the legend, preparatory to the rescan.

The slow manner in which lysozyme refolds when it is contained in a lysozyme aqueous solution 4.39 M in Gdm+ charge-compensated by an equimolar anion mixture (formate/acetate) is displayed in Figure 2. Each curve is obtained using a fresh sample from the same batch, which is then denatured by temperature increase to 75ºC and then immediately recooled at maximum instrumental rate (200K/min), to ambient temperature. After holding at ambient for a designated time, the sample is rescanned to 75ºC and the fraction that had refolded during the waiting time is assessed from the scan area compared with that of the initial unfolding. A similar series of scans is shown for the case in which the initial molarity of the mixed anion guanidinium solution is higher, 5.36M, and the refolding correspondingly slower, but eventually equally complete. For the faster complete refolding case, 4.39 M, we have also shown that the *same* sample can be repeatedly unfolded and allowed to refold and then unfolded again without change of unfolding enthalpy or $T_d$. Thus the system has acquired an important stability against aggregation, as well as ability to refold, even if slowly

The results for the different waiting times are plotted in Figure 4 for both Gdm+ concentrations, each case involving the same 10 wt % of dissolved lysozyme. It can be seen that the fraction of lysozyme that has refolded during the waiting time varies essentially linearly with log(waiting time), The refolding rate, as expected, decrease with increasing total [Gdm+]. It also decreases with increasing Fm/Ac ratio (not shown) and becomes incomplete at longer waiting times because of some aggregation process that we have not further explored.

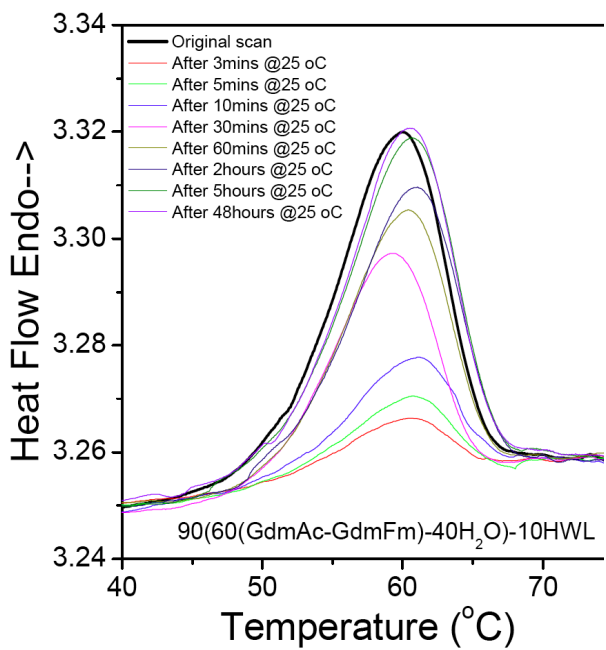
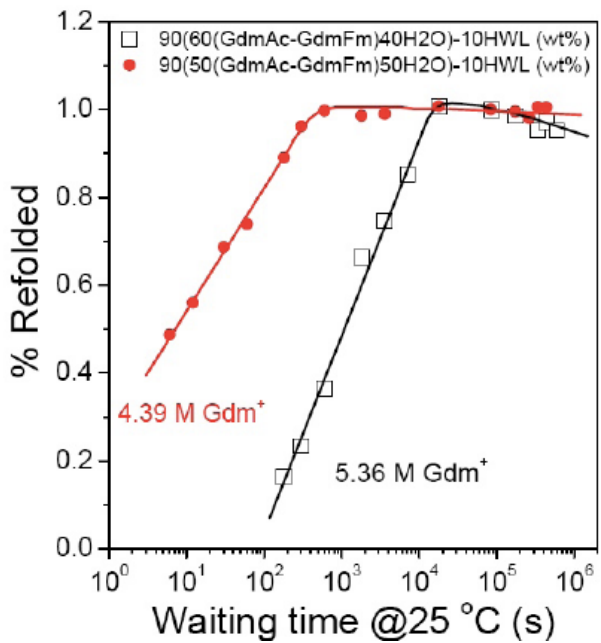

**Figure 3.** Unfolding endotherms for lysozyme in a GdmFm/GdmAc solution of Gdm+ concentration 5.36M. As for Fig. 2, dark curve is the endotherm for the initial unfolding scan and others are for scans after designated waiting times at ambient temperature .

**Figure 4.** logarithmic dependence of lysozyme refolding time for different solvent Gdm+ concentrations., 4.39 and 5.36M.=, as indicated in legend. The characteristic refolding times derived from these plots themselves increase exponentially with the concentration of Gdm+

We have also observed the effects of replacing a fraction of the Gdm+ cation with the similar-sized trimethylammonium cation (N(CH$_3$)$_3$H$^+$) and will report the findings elsewhere. The slopes of the plots and the fact that the fraction folded has a sharp, non-sigmoidal, arrival at 1.0, will be subject to analysis when more experimental data are at hand. Much characterization remains to be done, but we report here the important findings which we suggest point to essential conditions needed for refolding on short time scales.

The fact that the refolding rate is controlled by the Fm/Ac ratio is very significant. Formic acid is significantly stronger as a Bronsted acid than is acetic acid, which has a pKa value comparable to those of the aspartic and glutamic acid residues of the common proteins. It seems to us that in our experiments, by flooding the unfolded protein heteropolymer environment with acid groups of proton acitivity comparable to the protein carboxylates, we must be chemically pre-empting the formation of the electrostatic links that draw together distant parts of the unfolded heteropolymer to provide a critical preliminary configurations from which rapid downhill folding quickly determines arrival at the native structure. This picture is consistent with the knowledge(6) that, in lysozyme, the stability of the protein against misfolding to the lower free energy fibril state is compromised if Asp 67 is replaced by histidine, as occurs in mutants that predispose their carriers to systemic amylidosis(6). Likewise it is consistent with the studies of the Trp cage miniprotein by Garcia and co-workers(23) who identified the prior formation of the ionic link between Asp 9 and the arginine residue as the essential step that determined when the folding would occur. For this link, ΔpKa = 8.5 which we have found elsewhere(24) is quite sufficient to establish fully ionic behavior in protic ionic liquids (e.g. molten Gdm formate, T$_m$ = 91ºC(22)).

**Concluding Remarks.**

There are many possible variants of this anion tuning theme to examine. It will be interesting, for instance, to observe if the same slow folding behavior can be obtained with different combinations of anions, not necessarily carboxylate in character, that establish the same mean proton activity as in the present case. Likewise it will be important to find how specific the anion combination might be amongst different proteins, particularly in the extremophile cases.

**Acknowledgements**

This work was carried out under the auspices of the National Science Foundation Experimental Chemistry Division Collaborative Research Grant, No. CHE–0909120. Colleagues Pablo Debenedetti, Peter Rossky and H. Eugene Stanley, are thanked for stimulating discussions of the present systems and related "all-or-nothing" processes in physics and biology. Preliminary work on a mixed cation system containing Gdm chloride was carried out by Nolene Byrne (unpublished) and its features suggested the present study.

**Referencess**

1. Lim, W. K., J. Rösgen, S. W. Englandeer, Urea, but not guanidinium, destabilizes proteins by forming hydrogen bonds to the peptide group. *Proc. Nat. Acad. Sci.* **106**, 2595 (2009).
2. K. A. Dill, K. M. Fiebig, H. S. Chan, Cooperativity in protein folding kinetics. *Proc. Nat. Acad. Sci.* **90**, 1942 (1993).
3. A. R. Fersht, Characterizing transition states in protein folding: An essential step in the puzzle. . *Curr. Opin. Struct. Biol* **5**, 79 (1995.).
4. K. A. Dill, H. S. Chan, From Levinthal to pathways to funnels *Nature Structural Biology* **4**, 10 (1997).
5. T. R. Sosnick, D. Barrick, The folding of single domain proteins—have we reached a consensus? *Current Opinion in Structural Biology 2011, 21:12–24* **21**, 12 (2011).
6. D. R. Booth, M. Sunde, e. al, Instability, unfolding and aggregation of human lysozyme variants unerlying amyloid fibrilloogenesis. *Nature* **385**, 788 (1997).
7. S. E. Jackson, A. R. Fersht, Folding of chymotrypsin inhibitor 2.1. Evidence for a two state transition. *Biochemistry* **30**, 10428 (1991).
8. P. Jemth *et al.*, Demonstration of a low-energy on-pathway intermediate in a fast-folding protein by kinetics, protein engineering, and simulation. *Proc. Nat. Acad. Sci.* **101**, 6450 (2004).
9. N. D. Socci, J. N. Onuchic, Folding kinetics of protein-like heteropolymers. *J. Chem. Phys.* **101**, 1519 (1994).
10. V. I. Abkevich, A. M. Gutin, E. I. Shakhnovich, Specific nucleus as the transition state for protein folding: evidence from the lattice model. *Biochemistry* **33**, 10026 (1994).
11. H. S. Chan, K. A. Dill, Protein folding in the landscape perspective: Chevron plots and non-Arrhenius kinetics. *Protein Structure Function and Bioinformatics* **30**, 2 (1998).
12. D. Thirumalai, Z. Guo, Nucleation mechanism for proteifolding and theoretical predictions for hydrogen-exchange labeling experiments. *Biopolymers* **35**, 137 (1995).
13. K. A. Dill, Dominant Forces in Protein Folding. *Biochemistry* **29**, 7133 (1990).
14. V. Munoz, P. A. Thompson, J. Hofrichter, Folding dynamics and mechanism of beta-hairpin formation. *Nature* **390**, 196 (1997).
15. W. A. Eaton *et al.*, Fast kinetics and mechanisms in protein folding *Annual Review of Biophysics and Biomolecular Structure* **29**, 327 (2000).
16. P. Privalov, N. N. Khechinashvili, A thermodynamic approach to the problem of stabilization of globular protein structure: a calorimetric study. *J. Mol. Biol.* **86**, 665 (1974).
17. W. Pfeil, P. L. Privalov, Thermodynamic investigations of proteins: 1 standard functions for proteins with lysozyme as an example. *Biophys. Chem.* **4**, 23 (1976).
18. N. Byrne, C. A. Angell, Protein unfolding, and the "tuning in" of reversible intermediate states, in protic ionic liquid media. *J. Mol. Biol.* **378**, 707 (2008).

19. N. Byrne, J.-P. Belieres, C. A. Angell, Directed destabilization of lysozyme in protic ionic liquids reveals a compact, low energy, soluble, reversibly-unfolding (pre-fibril) state. *arxiv0710.3807*, (2007).
20. C. A. Angell, L.-M. Wang, Hyperquenching and cold equilibration strategies for the study of liquid–liquid and protein folding transitions. *Biophysical Journal* **105**, 621 (2003).
21. P. von Hippel, T. Schleich, *Biological Macromolecules Eds Timasheff S, Fasman, G.* **2**, (1969).
22. Z.-F. Zhao, K. Ueno, C. A. Angell, High Conductivity, and "Dry" Proton Motion, in Guanidinium Salt Melts and Binary Solutions. *J. Phys Chem. B (online)*, dx.doi.org/10.1021/jp206491z (2011).
23. C. A. Jimenez-Cruz, G. I. Makhatadze, A. E. Garcia, Protonation/deprotonation effects on the stability of the Trp-cageminiprotein. *Phys. Chem. Chem. Phys.* **13**, 17056 (2011).
24. M. Yoshizawa, W. Xu, C. A. Angell, Ionic Liquids by Proton Transfer: Vapor Pressure, Conductivity, and the Relevance of ΔpK$_a$ from Aqueous Solutions. *J. Amer. Chem. Soc.* **125**, 15411 (2003).